\documentclass[aps,reprint,superscriptaddress,longbibliography,pra]{revtex4-1}
\usepackage[percent]{overpic}
\usepackage{graphicx, subfigure}
\usepackage{bm}
\usepackage{times}
\usepackage{amsfonts}
\newcommand\lsc[1]{{}^{#1}}

\begin{document}

\title{Scattering in Multilayered Structures:\\Diffraction from a Nanohole.}

\author{Ivan \surname{Fernandez-Corbaton}\email{ivan.fernandez-corbaton@mq.edu.au}}
\affiliation{QISS, Department of Physics and Astronomy, Macquarie University, NSW 2109, Australia}
\affiliation{ARC Centre for Engineered Quantum Systems}
\author{Nora Tischler\email{nora.tischler@connect.qut.edu.au}}
\affiliation{QISS, Department of Physics and Astronomy, Macquarie University, NSW 2109, Australia}
\affiliation{Applied Optics and Nanotechnology Group, Discipline of Physics, Queensland University of Technology, QLD 4001, Australia}
\author{Gabriel Molina-Terriza\email{gabriel.molina-terriz@mq.edu.au}}
\affiliation{QISS, Department of Physics and Astronomy, Macquarie University, NSW 2109, Australia}
\affiliation{ARC Centre for Engineered Quantum Systems}

\begin{abstract}
The spectral expansion of the Green's tensor for a planar multilayered structure allows us to semi analytically obtain the angular spectrum representation of the field scattered by an arbitrary dielectric perturbation present in the structure. In this paper we present a method to find the expansion coefficients of the scattered field, given that the electric field inside the perturbation is available. The method uses a complete set of orthogonal vector wave functions to solve the structure's vector wave equation. In the two semi-infinite bottom and top media, those vector wave functions coincide with the plane-wave basis vectors, including both propagating and evanescent components. The technique is used to obtain the complete angular spectrum of the field scattered by a nanohole in a metallic film under Gaussian illumination. We also show how the obtained formalism can easily be extended to spherically and cylindrically multilayered media. In those cases, the expansion coefficients would multiply the spherical and cylindrical vector wave functions.  
\end{abstract}
\maketitle
\section{Introduction}

With the last decade's revolution in microscopy and lithographic techniques, the field of nanophotonics has experienced immense growth. This expansion has prompted the development of new numerical and analytical techniques for the electromagnetic study of nanostructures. Some of the most important theoretical methods in nanophotonics are reviewed in \cite[chap. 15]{Novotny2006}. The theoretical understanding of the interaction of light and matter at the nanoscale has allowed new discoveries in the fields of plasmonics, optical microscopy, and Raman scattering \cite{Weeber1999,Hetch2000,Nie1997}. A particular problem in nanophotonics, which has particularly benefited from the new theoretical techniques, has been the extraordinary optical transmission of light discovered by Ebbessen {\em et al} \cite{Ebbesen1998}. This effect appears when we scatter light off back-illuminated subwavelength hole arrays in metallic thin films and has found applications in sensing and lens design (\cite{Brolo2004,Huang2008}). Since then, several authors have analyzed the transmission of light through isolated nanoholes in metallic thin films and the collective behavior of arrays of nanoholes (see the extensive review in \cite{Garcia-Vidal2010} and references therein).

In this work we present a technique which allows the plane-wave decomposition of the field scattered by an arbitrary perturbation present in a planar multilayered structure under general external excitation to be obtained. The technique is applied to the problem of a single back-illuminated nanohole in a metallic film. We show the exact angular spectrum representation of the scattered electric field containing both propagating and evanescent components. To the best of our knowledge, this detailed field decomposition has not yet been studied for these nanostructures, even though the problem has received a significant amount of attention in recent years (for example, Refs. \cite{Lalanne2005,Sepulveda2008,Nikitin2010,Johansson2010,Bordo2011}). Understanding the spatial dependence of the scattered fields is necessary to efficiently collect the light transmitted through the nanoapertures, but more importantly, it can provide important information about the electromagnetic response of the nanostructure, which can then be used in applications like photovoltaic design \cite{Ferry2008}, optical trapping \cite{Kwak2004}, and nano-optical tweezers \cite{Juan2011}.

We finish the paper by outlining how the technique can be extended to spherically and cylindrically multilayered systems to obtain scattered field decompositions in spherical and cylindrical vector wave functions.

\section{Derivation of the method}
\label{sec:method}
A monochromatic $\exp(-iwt)$ time dependence of the field has been assumed in this analysis.
When faced with an electromagnetic scattering problem, the dyadic Green's tensor (GT) technique can be used to solve it \cite{Martin1998}, offering some advantages over other tools like versatility and ease of calculation. Let us consider a fairly general scattering problem where a nonmagnetic base system, characterized by a space-dependent relative dielectric constant $\varepsilon_b(\mathbf{r})$, is illuminated by a general excitation $\mathbf{E_0}(\mathbf{r})$. Now, in some region of space $V$, we introduce a perturbation, i.e., the points $\mathbf{r'}\in V$ have a different dielectric constant with respect to the unperturbed system:
\begin{equation}
\varepsilon(\mathbf{r'})=\varepsilon_b(\mathbf{r'})+\Delta\varepsilon(\mathbf{r'}).
\end{equation}
Martin and Piller showed in \cite{Martin1998} that the total electric field $\mathbf{E}(\mathbf{r})$ in the modified system can be expressed as:
\begin{equation}
\label{eq:dyson}
\mathbf{E}=\mathbf{E_0}+\int_{V}k_0^2\Delta\varepsilon(\mathbf{r'})\stackrel{\leftrightarrow}{\mathbf{G}}(\mathbf{r},\mathbf{r'})\mathbf{E}(\mathbf{r'})d\mathbf{r'},
\end{equation}
where 
\begin{itemize}
\item $\mathbf{E_0}(\mathbf{r})$ is the electric field due to the external excitation that would be present in $r$ if the perturbation did not exist, 
\item $k_0^2$ is the square of the vacuum wavenumber,
\item $\stackrel{\leftrightarrow}{\mathbf{G}}(\mathbf{r},\mathbf{r'})$ is the GT of the base (unperturbed) system, and
\item $\mathbf{E}(\mathbf{r'})$ is the total electric field present inside the perturbation.
\end{itemize}
Clearly, the integral of the right-hand side of Eq. (\ref{eq:dyson}) corresponds to the field scattered by the perturbation.
One of the difficulties in applying this method is that obtaining the GT for an arbitrary base system is a daunting task. Fortunately, when the system of interest has strong symmetries like stratification, its GT can be calculated. References \cite{Barkeshli1992}, \cite{Li1994} and \cite{Xiang1996} provide the GT for planarly, spherically and cylindrically multilayered structures, respectively. In an excellent contribution, Tan and Tan \cite{Tan1998} developed a unified formulation for the three mentioned geometries which is also valid for biisotropic media. In all cases, the construction of the GT is based upon its expansion in the kind of vector wave functions appropriate for each geometry: rectangular, spherical or cylindrical vector wave functions \cite[sec. 4]{Tan1998}. In this expansion, the GT is expressed as a sum of products of functions. Each term in the sum consists of the product of two functions, one which depends on $r$ and the other on $\mathbf{r'}$. This separate dependence is the key property exploited by our technique.
\begin{figure}[tbp]
  \begin{center}
	\includegraphics[width=8cm]{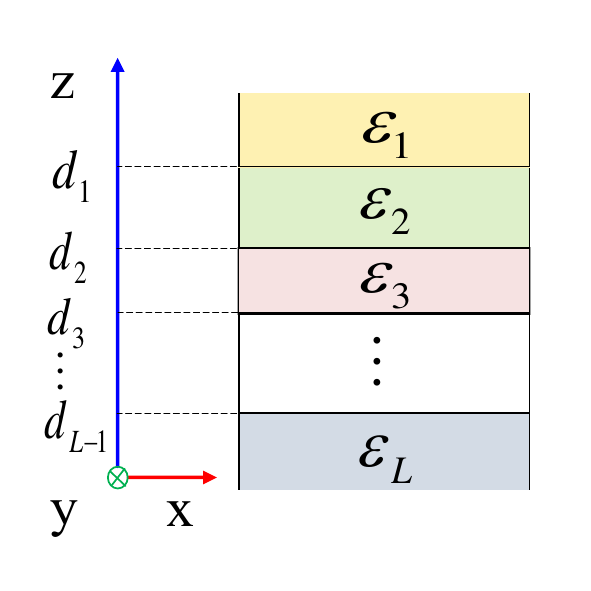}
   \end{center}
\caption{\label{fig:multilayer}(Color online) Geometry of the unperturbed system where the Green tensor is calculated. We consider a nonmagnetic planar multilayered structure, which is homogeneous in the $(x,y)$ plane. The positions of the interfaces are given by $d_l$ and the relative dielectric constant of each slab is given by $\varepsilon_l$.}
\end{figure}

\subsection{Green's tensor for a Planar Multilayered Structure}
In this section we obtain a modal decomposition of the GT for a planar multilayered structure. The dyadic GT of a system represents the system's electromagnetic response to an infinitesimal excitation and allows computation of the electric field induced in the system by an arbitrary distribution of electromagnetic sources \cite[chap 2.10]{Novotny2006}. Let us examine the planar multilayered structure depicted in Fig. \ref{fig:multilayer}. Each layer $l\in 1\ldots L$ has a different relative dielectric constant $\varepsilon_l$ and relative magnetic permeability equal to one. The planar boundaries between layers are perpendicularly oriented to the $\mathbf{\hat{z}}$ direction and they are located at $z$ coordinates $d_l,\ l\in 1\ldots L-1$. The layers extend to infinity in the $x$ and $y$ dimensions.

Using \cite{Barkeshli1992} or \cite[sec. 6]{Tan1998}, it is possible to calculate the effect of an electric dipole point source located at point $\mathbf{r'}=[x^\prime,y^\prime,z^\prime]$ in a given layer $l'$ on a target point $\mathbf{r}=[x,y,z]$ in layer $l$. Then we can arrive at the spectral decomposition of the structure's GT :
\begin{widetext}
\begin{eqnarray}
\label{eq:GT}
\nonumber
\stackrel{\leftrightarrow}{\mathbf{G}}(\mathbf{r},\mathbf{r^\prime})&=&-\frac{\mathbf{\hat{z}}\otimes\mathbf{\hat{z}}}{k_{l'}^2}\delta(\mathbf{r}-\mathbf{r'})\\
             &+& \int_{-\infty}^{+\infty}\int_{-\infty}^{+\infty}dk_x  dk_y \,\left[ \lsc{M}\alpha_{l'l,k_xk_y}^{\pm} (\mathbf{M}_{k_xk_y}^{\pm}(\mathbf{r})\otimes \mathbf{M}_{-k_x-k_y}^{\mp}(\mathbf{r'}))+\lsc{N}\alpha_{l'l,k_xk_y}^{\pm} (\mathbf{N}_{k_xk_y}^{\pm}(\mathbf{r})\otimes \mathbf{N}_{-k_x-k_y}^{\mp}(\mathbf{r'})) \right]
\end{eqnarray}
\begin{eqnarray}
\label{eq:M}
\mathbf{M}_{k_xk_y}^{\pm}(\mathbf{r})&=&\ i(\exp(\pm i k_z^l z)+\lsc{s}R_{l,k_xk_y}^{\pm}\exp(\mp i k_z^l z))\mathbf{\hat{s}}_{l,k_xk_y}\,\exp(i(k_xx+k_yy))\\
\label{eq:N}
\mathbf{N}_{k_xk_y}^{\pm}(\mathbf{r})&=&(\exp(\pm i k_z^l z)\mathbf{\hat{p}}_{l,k_xk_y}^{\pm}+\lsc{p}R_{l,k_xk_y}^{\pm}\exp(\mp i k_z^l z)\mathbf{\hat{p}}_{l,k_xk_y}^{\mp})\,\exp(i(k_xx+k_yy)),
\end{eqnarray}
\end{widetext}

where $\mathbf{\hat{a}}\otimes\mathbf{\hat{b}}$ is the outer product of the three-dimensional vectors $\mathbf{\hat{a}}$ and $\mathbf{\hat{b}}$; $k_l=k_0\sqrt{\varepsilon_l}$ are the wave numbers for $l=1\ldots L$; $[k_x,k_y]$ are the transversal components of the wavevector, which are assumed to take only real values throughout the paper; $\lsc{v}\alpha_{l'l,k_xk_y}^{\pm}$ for $v=\{M,N\}$ are complex scalars which depend on the transversal wavevector, the source and target layer indexes, and the type of vector wave function ($\mathbf{M}$ or $\mathbf{N}$); the rectangular vector wave functions $\mathbf{M}_{k_xk_y}^{\pm}(\mathbf{r})$ and  $\mathbf{N}_{k_xk_y}^{\pm}(\mathbf{r})$ are the mode functions; \footnote{They are not the same $\mathbf{M}$ and $\mathbf{N}$ rectangular vector wave functions defined in \cite[sec. 4.1]{Tan1998}. In our notation $\mathbf{M}$ and $\mathbf{N}$ already contain upward and downward propagating components. This allows more compact expressions.} and for the $\pm$ selection, the signs on top are to be taken when $z\ge z'$ and the bottom signs when $z<z'$.

In Eqs. (\ref{eq:M}) and (\ref{eq:N}), $k_z^l=\sqrt{k_l^2-(k_x^2+k_y^2)}$ where the branch with $\text{Im}\{k_z^l\}>0$ is chosen, and the definitions of the polarization vectors as a function of the transversal wavevector and the layer are, in the Cartesian basis,
\begin{eqnarray}
\label{eq:s}
\mathbf{\hat{s}}_{l,k_xk_y}&=&\frac{(k_y \mathbf{\hat{x}} - k_x \mathbf{\hat{y}})}{\sqrt{k_x^2+k_y^2}},\\
\label{eq:p}
\mathbf{\hat{p}}_{l,k_xk_y}^{\pm}&=&\frac{(\pm k_z^l (k_x \mathbf{\hat{x}} + k_y \mathbf{\hat{y}}) -(k_x^2+k_y^2)\mathbf{\hat{z}})}{\sqrt{k_x^2+k_y^2}\ k_l}.
\end{eqnarray}
These expressions coincide with the definitions of the S and P polarization vectors of a plane wave with transversal wavevector $[k_x,k_y]$ propagating in a multilayered structure \cite{Paulus2000}.

Finally, we give the expressions for the expansion coefficients:
\begin{equation}
\label{eq:alpha}
\lsc{v}\alpha_{l'l,k_xk_y}^{\pm}=\frac{i}{8\pi k_z^{l'}}\frac{\lsc{v}T_{l'l,k_xk_y}^{\pm}}{1-\lsc{v}R_{l',k_xk_y}^-\lsc{v}R_{l',k_xk_y}^+},
\end{equation}
for $v=\{M,N\}$, or equivalently $v=\{s,p\}$.

The $\lsc{v}R_{l,k_xk_y}^{\pm}$ and $\lsc{v}T_{l'l,k_xk_y}^{\pm}$ quantities are sometimes referred to as generalized Fresnel reflection coefficients and generalized transmission coefficients (see \cite[chap. 3.3]{Kong1990}). Recursive methods for computing $\lsc{v}R_{l,k_xk_y}^{\pm}$ and $\lsc{v}T_{l'l,k_xk_y}^{\pm}$ are detailed in \cite{Barkeshli1992}, \cite{Tan1998}, \cite{Paulus2000},  and \cite[chap. 3.3]{Kong1990}.

Similarly to \cite{Li2004}, we can establish orthogonality relationships between the vector wave functions. Given the following scalar product between two vectorial fields: 
\begin{equation}
\label{eq:sprod}
\langle \mathbf{V}|\mathbf{W} \rangle = \int_{-\infty}^{+\infty}\int_{-\infty}^{+\infty}\int_{-\infty}^{+\infty} d\mathbf{r} \bar{\mathbf{V}}(\mathbf{r})\mathbf{W}(\mathbf{r}),
\end{equation}
where $\bar{\mathbf{V}}(\mathbf{r})$ is the Hermitian conjugate of $\mathbf{V}(\mathbf{r})$, and it is easy to verify that $\langle \mathbf{M}_{k_x,k_y}^+|\mathbf{M}_{q_x,q_y}^+\rangle=0$ unless $[k_x,k_y]=[q_x,q_y]$; $\langle \mathbf{N}_{k_x,k_y}^+|\mathbf{N}_{q_x,q_y}^+\rangle=0$ unless $[k_x,k_y]=[q_x,q_y]$; and $\langle \mathbf{M}_{k_x,k_y}^+|\mathbf{N}_{q_x,q_y}^+\rangle=0$ for all $[k_x,k_y]$ and $[q_x,q_y]$.

The same orthogonality relationships apply for the minus superscripted vector wave functions.

From now on, when applying (\ref{eq:GT}) to the GT between a target point $\mathbf{r}=[x,y,z]$ and a set of source points $\mathbf{r'}=[x',y',z']\in V_J$ we will make the restriction 
\begin{equation}
\text{either } z>z'\ \forall(z,z')\text{ or } z<z'\ \forall(z,z').
\label{eq:ass}
\end{equation}
As we will show shortly, this allows us to write our expressions in a very compact form. Also, for most scattering problems, the positions of electromagnetic sources and target points satisfy (\ref{eq:ass}). In the next section this restriction will be applied to the positions of scatterers and target points as well. 

Taking (\ref{eq:ass}) into account, we simplify the notation defining:
\begin{eqnarray*}
\mathbf{e}_{k_xk_y}^s(\mathbf{r})&=&\mathbf{M}_{k_xk_y}(\mathbf{r})\  ,\  \mathbf{e}_{k_xk_y}^p(\mathbf{r})=\mathbf{N}_{k_xk_y}(\mathbf{r})\\
\mathbf{f}_{k_xk_y}^s(\mathbf{r'})&=&\mathbf{M}_{-k_x-k_y}(\mathbf{r'})\ ,\ \mathbf{f}_{k_xk_y}^p(\mathbf{r'})=\mathbf{N}_{-k_x-k_y}(\mathbf{r'}),\\
\end{eqnarray*}
where we drop the $\pm$ sign, which clutters the expressions and can easily be recovered with respect to (\ref{eq:GT}) depending on $sign(z-z')$. We now re-write (\ref{eq:GT}) with the newly defined vector fields:
\begin{eqnarray}
\label{eq:GT2}
&&\stackrel{\leftrightarrow}{\mathbf{G}}(\mathbf{r},\mathbf{r'})=-\frac{\mathbf{\hat{z}}\otimes\mathbf{\hat{z}}}{k_{l'}^2}\delta(\mathbf{r}-\mathbf{r'})\\
\nonumber
&&+\sum_{v=s,p}\int_{-\infty}^{+\infty} dk_x \int_{-\infty}^{+\infty} dk_y \alpha_{l'l,k_xk_y}^v \mathbf{e}_{k_xk_y}^v(\mathbf{r})\otimes \mathbf{f}_{k_xk_y}^v(\mathbf{r'}).
\end{eqnarray}

When the problem setup complies with restriction (\ref{eq:ass}), we can use Eq. (\ref{eq:GT2}) to compute the field in a point $r$ of the unperturbed base system due to a source distribution $\mathbf{J}(\mathbf{r})$ present in volume $V_J$:
\begin{widetext}
\begin{eqnarray}
\mathbf{E_0}(\mathbf{r})&=&\int_{V_J} d\mathbf{r'}\stackrel{\leftrightarrow}{\mathbf{G}}(\mathbf{r},\mathbf{r'})\cdot \mathbf{J}(\mathbf{r'}) \nonumber\\
&=&\int_{V_J}d\mathbf{r'}\big[-\frac{\mathbf{\hat{z}}\otimes\mathbf{\hat{z}}}{k_{l'}^2}\delta(\mathbf{r}-\mathbf{r'})+\sum_{v=s,p}\int_{-\infty}^{+\infty} dk_x \int_{-\infty}^{+\infty} dk_y \alpha_{l'l,k_xk_y}^v \mathbf{e}_{k_xk_y}^v(\mathbf{r})\otimes \mathbf{f}_{k_xk_y}^v(\mathbf{r'})\big]\cdot \mathbf{J}(\mathbf{r'}) \nonumber\\
&=&-\frac{\mathbf{\hat{z}}\otimes\mathbf{\hat{z}}}{k_{l'}^2}\mathbf{J}(\mathbf{r})+\sum_{v=s,p}\int_{-\infty}^{+\infty} dk_x \int_{-\infty}^{+\infty} dk_y \mathbf{e}_{k_xk_y}^v(\mathbf{r}) (\int_{V_J} d\mathbf{r'}\alpha_{l'l,k_xk_y}^v \mathbf{f}_{k_xk_y}^v(\mathbf{r'})\cdot \mathbf{J}(\mathbf{r'})),\nonumber\\
\label{eq:almost}
&=&-\frac{\mathbf{\hat{z}}\otimes\mathbf{\hat{z}}}{k_{l'}^2}\mathbf{J}(\mathbf{r})+\sum_{v=s,p}\int_{-\infty}^{+\infty} dk_x \int_{-\infty}^{+\infty} dk_y \mathbf{e}_{k_xk_y}^v(\mathbf{r})\beta_{l'l,k_xk_y}^v.
\end{eqnarray}
\end{widetext}
The first equality follows from using the GT dyadic $\stackrel{\leftrightarrow}{\mathbf{G}}(\mathbf{r},\mathbf{r'})$ to solve the inhomogeneous vector wave equation of the multilayered structure. As indicated in \cite[chap. 13.1]{Morse1953}, the GT can be used to solve such a system within a boundary surface $\mathcal{S}$ where the field must meet the required boundary conditions. The first line of Eq. (\ref{eq:almost}) is obtained for $\mathcal{S}$ tending to infinity and enforcing that the field is zero at surface $\mathcal{S}$.

Outside volume $V_J$, where $\mathbf{J}(\mathbf{r})$ vanishes, the field is a weighted sum of the basis functions $\mathbf{e}_{k_xk_y}^v(\mathbf{r})$. The complex weight is given by:
\begin{equation}
\beta_{l'l,k_xk_y}^v=\int_{V_J}d\mathbf{r'}\alpha_{l'l,k_xk_y}^v \mathbf{f}_{k_xk_y}^v(\mathbf{r'})\cdot \mathbf{J}(\mathbf{r'}).
\end{equation}
The operation $a\cdot b$ represents either the product of a matrix and a vector or the inner product of two vectors. 
From Eq. (\ref{eq:almost}) we observe that, if we exclude the source points, the set of vector wave functions $\mathbf{e}_{k_xk_y}^v(\mathbf{r})$ is a complete basis for the field distributions in the base structure. Since we have already established their orthogonality, we conclude that, except at the source points,  the set of vector wave functions $\mathbf{e}_{k_xk_y}^v(\mathbf{r})$ is a complete orthogonal basis for the field distributions in the structure. Recall that depending on which side of restriction (\ref{eq:ass}) is met, $\mathbf{e}_{k_xk_y}^v(\mathbf{r})=\{\mathbf{M}_{k_x,k_y}^+(\mathbf{r}),\mathbf{N}_{k_x,k_y}^+(\mathbf{r})\}$ or $\mathbf{e}_{k_xk_y}^v(\mathbf{r})=\{\mathbf{M}_{k_x,k_y}^-(\mathbf{r}),\mathbf{N}_{k_x,k_y}^-(\mathbf{r})\}$.

We can now comment on the convenience of using restriction (\ref{eq:ass}). It is clear that without imposing it we would get a spectral expansion with terms in both basis $\{\mathbf{M}_{k_x,k_y}^+(\mathbf{r}),\mathbf{N}_{k_x,k_y}^+(\mathbf{r})\}$ and $\{\mathbf{M}_{k_x,k_y}^-(\mathbf{r}),\mathbf{N}_{k_x,k_y}^-(\mathbf{r})\}$, which would impair the clarity of the results and would be cumbersome to use.

\subsection{Dielectric Perturbations}
Now we are ready to get to the core of our technique. Let us go back to Eq. (\ref{eq:dyson}) describing the scattering problem when a perturbation with volume $V$ and offset dielectric constant $\Delta\varepsilon(\mathbf{r})$ is introduced in the multilayered structure. The solution of that equation is found in two steps:
\begin{enumerate}
\item By discretizing the perturbation volume $V$ in $T$ small volumes with centers at $\mathbf{r'}=\mathbf{r_t}$, and setting $\mathbf{r}=\mathbf{r_t}\ \forall \ t \in \{1\ldots T\}$, a self-consistent $3T\times3T$ system of equations arises for the field inside the perturbation $\mathbf{E}(\mathbf{r_t})$.
\item Once $\mathbf{E}(\mathbf{r_t})$ is available, the scattered field $\mathbf{E_{sc}}(\mathbf{r})$ [integral term of Eq. (\ref{eq:dyson})] and the total field $\mathbf{E}(\mathbf{r})$ can be computed for $r\notin V$ using $\mathbf{E}(\mathbf{r'})$ as the generator in Eq. (\ref{eq:dyson}). The volume integral becomes a finite sum over each discrete volume element. We keep the continuous notation for convenience.
\end{enumerate}
The first step is by no means trivial: see the introduction in \cite{Piller1998} for a summary of the encountered difficulties. There are several possibilities to compute $\mathbf{E}(\mathbf{r'})$ (\cite[chap. 15]{Novotny2006}), but all of them typically have problems of convergence when the contrast between the scatterer and the medium is high. In this paper we are interested in the expansion of the scattered field, so we do not discuss the subtleties of calculating the internal field $\mathbf{E}(\mathbf{r'})$ for $\mathbf{r'}\in V$. For all purposes of this paper the internal field can be considered as an input to our problem. Just for the sake of completeness we mention that we calculated this field by using the self-consistent  method given in \cite{Paulus2001}, but note that our technique is totally independent of the particular method employed to calculate $\mathbf{E}(\mathbf{r'})$.

We now focus on the integral part of the right hand side of Eq. (\ref{eq:dyson}) when $r\notin V$. As mentioned, this is the term that corresponds to the field scattered by the perturbation. Noting that $\mathbf{e}_{k_xk_y}^v(\mathbf{r})$ does not depend on the integration variable $\mathbf{r'}$ and using the decomposition of $\stackrel{\leftrightarrow}{\mathbf{G}}(\mathbf{r},\mathbf{r'})$ for $\mathbf{r}\neq \mathbf{r'}$ in (\ref{eq:GT2}) and the associativity of the vector outer and inner products, we arrive at our main result. By expressing the volume integral in (\ref{eq:dyson}) as

\begin{eqnarray}
\label{eq:mainresult}
\mathbf{E_{sc}}(\mathbf{r})&=&\sum_{v=s,p}\int_{-\infty}^{+\infty} \int_{-\infty}^{+\infty} dk_x dk_y\, \gamma_{k_xk_y}^v \mathbf{e}_{k_xk_y}^v(\mathbf{r}),\\
\label{eq:gamma}
\gamma_{k_xk_y}^v&=&\int_Vd\mathbf{r'}\alpha_{l'l,k_xk_y}^v \mathbf{f}_{k_xk_y}^v(\mathbf{r'})\cdot \mathbf{E}(\mathbf{r'})k_0^2\Delta\varepsilon(\mathbf{r'}),
\end{eqnarray}
we have hence arrived at a decomposition for $\mathbf{E_{sc}}(\mathbf{r})$ in the vector basis modal functions of the unperturbed structure $\mathbf{e}_{k_xk_y}^v(\mathbf{r})$ $\forall \ r\notin V$. A restriction of the type (\ref{eq:ass}) has been assumed in (\ref{eq:mainresult}). This time it applies to the positions of the scattering volume $V$ with respect to the the target point(s) $r$. This assumption is again true in many scenarios of interest.

We now turn our attention to the illuminating field $\mathbf{E_0}(\mathbf{r})$. Assuming that the target points we are interested in do not contain electromagnetic sources, Eq. (\ref{eq:almost}) means that $\mathbf{E_0}(\mathbf{r})$ can be expanded in the $\mathbf{e}_{k_xk_y}^v(\mathbf{r})$ basis by some complex scalars $\beta_{l'l,k_xk_y}^v$. For the illuminating field, the source layer is the one where the active light source is. We define it to be the first one $l'=1$:
\begin{equation}
\label{eq:E0}
\mathbf{E_0}(\mathbf{r})=\sum_{v=s,p}\int_{-\infty}^{+\infty} dk_x \int_{-\infty}^{+\infty} dk_y \beta_{1l,k_xk_y}^v \mathbf{e}_{k_xk_y}^v(\mathbf{r}).
\end{equation}

The entire system of equations is linear in the incident field $\mathbf{E_0}(\mathbf{r})$. This can be easily checked by formally rewriting Eq. (\ref{eq:dyson}) inside the perturbation volume $V$ as an operator equation, i.e. $\mathbf{E_0}(\mathbf{r'})=\mathcal{L}\mathbf{E}(\mathbf{r'})$. The integral operator $\mathcal{L}$ is linear. Thus, upon inversion the system remains linear, which means that the field inside the perturbation depends linearly on the incident field $\mathbf{E_0}(\mathbf{r})$. It follows that by computing the decomposition of the field scattered by the perturbation when illuminated by each of the $\mathbf{e}_{k_xk_y}^v(\mathbf{r})$ basis functions, we can obtain the decomposition of $\mathbf{E_{sc}}(\mathbf{r})$ due to an arbitrary excitation $\mathbf{E_0}(\mathbf{r})$ by mere linear superposition.

Note that in the top and bottom semi-infinite layers, and because there are no reflections from $\pm \infty$, $\lsc{v}R_{1,k_xk_y}^+=\lsc{v}R_{L,k_xk_y}^-=0\ \forall \ [v,k_x,k_y]$. This means that in those layers $\mathbf{e}_{1,k_xk_y}^v(\mathbf{r})$ and $\mathbf{e}_{L,k_xk_y}^v(\mathbf{r})$ are just plane-waves [see (\ref{eq:M}) and (\ref{eq:N})]. Note that both propagating ($\text{Im}\{k_z^{1,L}\}=0$) and evanescent ($\text{Im}\{k_z^{1,L}\}\neq0$) plane waves are present in the expansion, giving us all the information contained in $\mathbf{E_{sc}}(\mathbf{r})$. 

On those two layers, the type of expansion in equations (\ref{eq:mainresult}) and (\ref{eq:E0}) is commonly known as the angular representation of a field: \cite[chap 2.12]{Novotny2006} and \cite[chap 3.2]{Mandel1995}.

Let us now summarize the importance of results (\ref{eq:mainresult}-\ref{eq:E0}). We formally start with an electromagnetic source $\mathbf{J}(\mathbf{r})$, away from the perturbation, which induces the field $\mathbf{E_0}(\mathbf{r})$ in the structure. This field is a weighted sum of the $\mathbf{e}_{k_xk_y}^v(\mathbf{r})$ functions. When the field $\mathbf{E_0}(\mathbf{r})$ encounters the perturbation $\Delta\varepsilon(\mathbf{r'})$ in $\mathbf{r'}\in V$ it interacts with it and, as a result, a scattered field $\mathbf{E_{sc}}(\mathbf{r})$ is added to the original $\mathbf{E_0}(\mathbf{r})$. This new field is again a weighted sum of the $\mathbf{e}_{k_xk_y}^v(\mathbf{r})$ functions when $r\notin V$. 

In less abstract terms, assume for instance that an S (or P) linearly polarized plane wave illuminates an unperturbed planar multilayered structure. The $[k_x,k_y]$ components of the plane wave and its polarization will correspond to one of the $\mathbf{e}_{k_xk_y}^v(\mathbf{r})$, which will exist in the structure without ever changing $[k_x,k_y]$ or $v$. On the other hand, if there is a perturbation in the structure, the original plane wave will give rise to an infinite number of new $[v,k_x,k_y]$ components with weights given by (\ref{eq:gamma}). 

\section{Application of the Method to a Nanohole}
\label{sec:results}
We envision many possible uses for this technique. In this section we will apply it to the well known problem of a nanohole in a metallic film illuminated by a Gaussian beam. In this way, we will show how to use this method in a typical problem and, at the same time, we will calculate the shape of the scattered field from a nanohole, which may be of importance for certain applications.

\begin{figure}[htbp]
  \begin{center}
	\includegraphics[width=8cm]{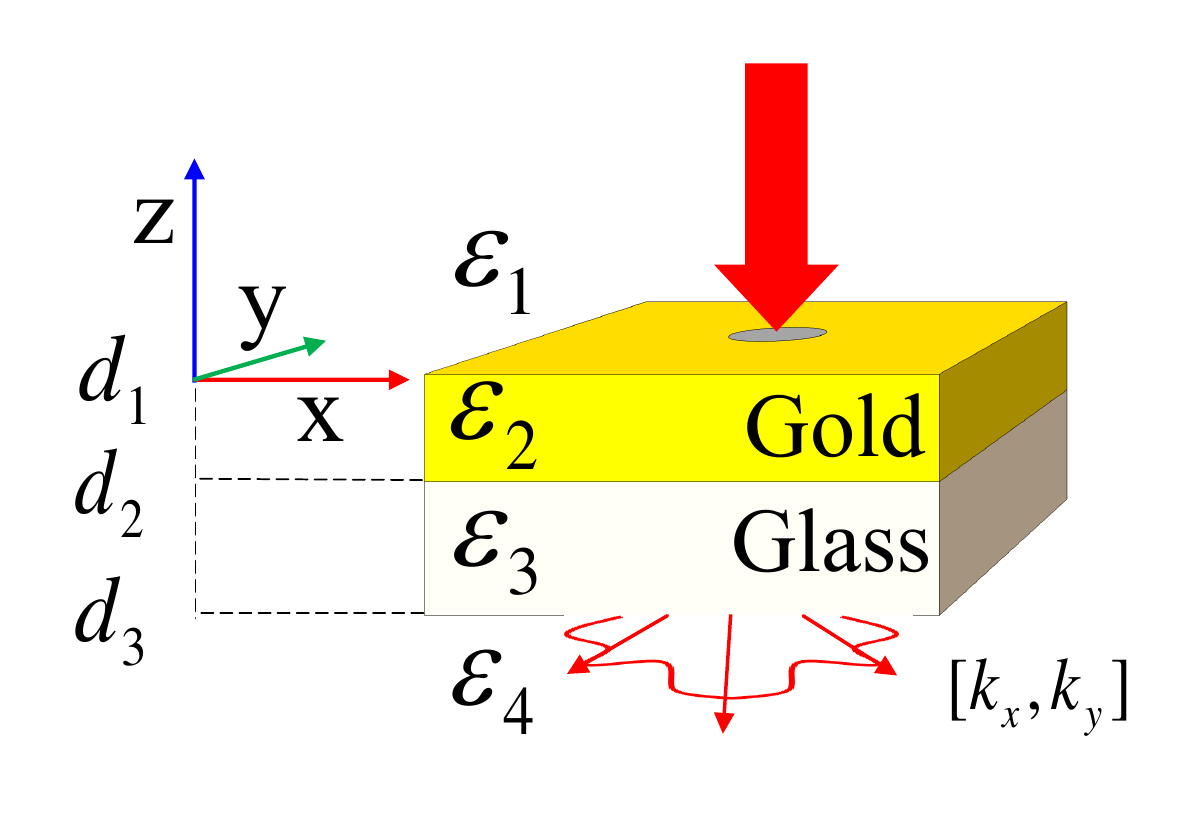}
   \end{center}
\caption{\label{fig:nanohole}(Color online) Setup for nanohole scattering analysis. The excitation is a circularly polarized Gaussian beam propagating through the air side and impinging on the gold layer with zero incidence angle. The excitation wavelength is 950 nm and its Gaussian beam waist 1.5 $\mu$m. The sample consists of a 400-nm-diameter nanohole made on a 140-nm layer of gold deposited on a 200-nm glass substrate. In the figure $\varepsilon_1=\varepsilon_4=1$, $\varepsilon_2=\varepsilon_{Au}=-37.2+2.5i$, and $\varepsilon_3=\varepsilon_{glass}=2.25$. Also $d_1=0$, $d_2=-140$ nm, and $d_3=-340$ nm. }
\end{figure}

Fig. \ref{fig:nanohole} shows a sketch of the considered system. The angular spectrum decomposition is calculated at the air layer opposite of where the excitation comes from. The laser layer is the first one, the perturbation is contained in the gold layer $l'=2$ and the target layer is $l=4$. Here is a step by step description of how the method is applied.  

To represent the Gaussian illumination, a set of basis functions $\mathbf{e}_{k_xk_y}^v(\mathbf{r})$ is selected which needs to be sufficiently dense in $[k_x,k_y]$. We call it $\bm{I}$. Since all its components are propagating in vacuum, the set can be restricted to functions whose tranversal wavevector lies inside the circle $k_0^2=k_x^2+k_y^2$. We can then use the plane wave expansion of a Gaussian beam from \cite{Gabi2008} in order to compute the expansion coefficients $\beta_{11,k_xk_y}^v$ in equation (\ref{eq:E0}). The relationship to the expansion coefficients in other layers is:

\begin{equation}
\label{eq:beta}
\beta_{1l,k_xk_y}^v=\beta_{11,k_xk_y}^v \lsc{v}T_{1l,k_xk_y}^{-}.
\end{equation}

The interior of the nanohole is discretized at points $\mathbf{r_t}$. We assume that the field $\mathbf{E}_{k_xk_y}^v(\mathbf{r_t})$ present inside the perturbation when the external excitation is equal to each of the modes $\mathbf{e}_{k_xk_y}^v(\mathbf{r})\in\bm{I}$ is available. In our case, we have used a self-consistent dipole method to obtain it.

We will define now the output basis set $\bm{O}$ of vectors $\mathbf{e}_{k_xk_y}^v(\mathbf{r})$ over which the angular spectrum will be computed. The basis vectors in $\bm{I}$ and in $\bm{O}$ can be different, in order to properly describe the different properties of the incoming beam and the scattered field. In particular, if we are interested in the evanescent as well as the propagating components of the scattered field, the set $\bm{O}$ has to be chosen accordingly. 

Let us now select one of the $\mathbf{E}_{k_xk_y}^v(\mathbf{r_t})$. Using (\ref{eq:gamma}) with the volume integral approximated by a discrete sum over the mesh points $\mathbf{r_t}$ we can obtain the expansion coefficients $\gamma_{k_xk_y}^v$ for each element of $\bm{O}$. At the target layer, this set of coefficients is the angular spectrum of the scattered field if the excitation was only the mode $\mathbf{e}_{k_xk_y}^v(\mathbf{r})\in\bm{I}$. This is repeated for every element in $\bm{I}$ and all the partial contributions are added with weights $\beta_{11,k_xk_y}^v$ to obtain the final result.

Should the ouput set $\bm{O}$ be a dense enough sampling of the $[k_x,k_y]$ space, a trivial integral in the momentum space would recover the value of the total electrical field $\mathbf{E_{sc}}(\mathbf{r})$ at any point $r$ of the structure meeting (\ref{eq:ass}) by just using (\ref{eq:mainresult}).

\begin{figure*}[htbp]
  \begin{center}
  \subfigure{%
            \label{fig1:S}
			\begin{overpic}[width=8cm,tics=10]{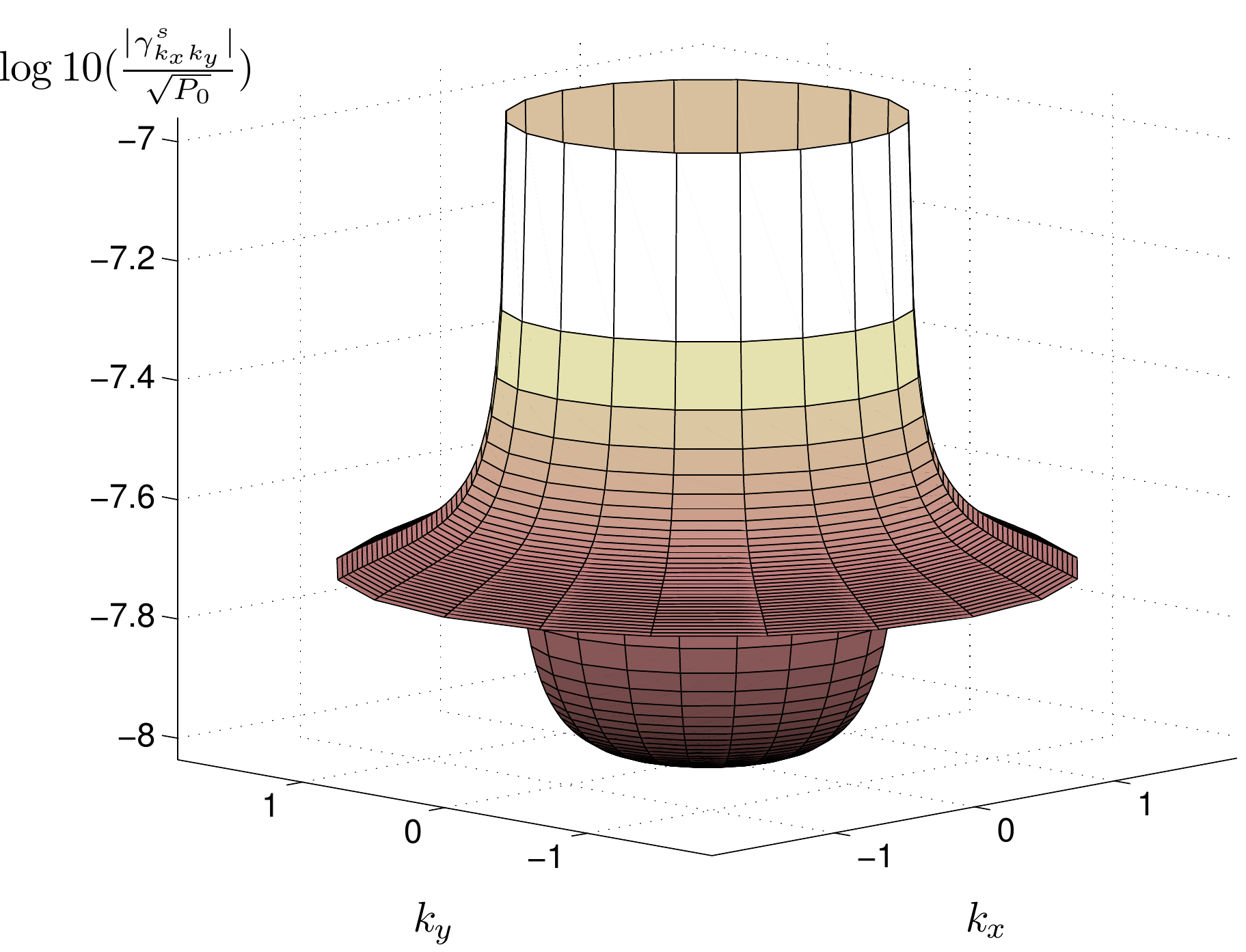}
			\put(90,70){\thesubfigure}
			\end{overpic}
			}
  \subfigure{%
            \label{fig1:P}
			\begin{overpic}[width=8cm,tics=10]{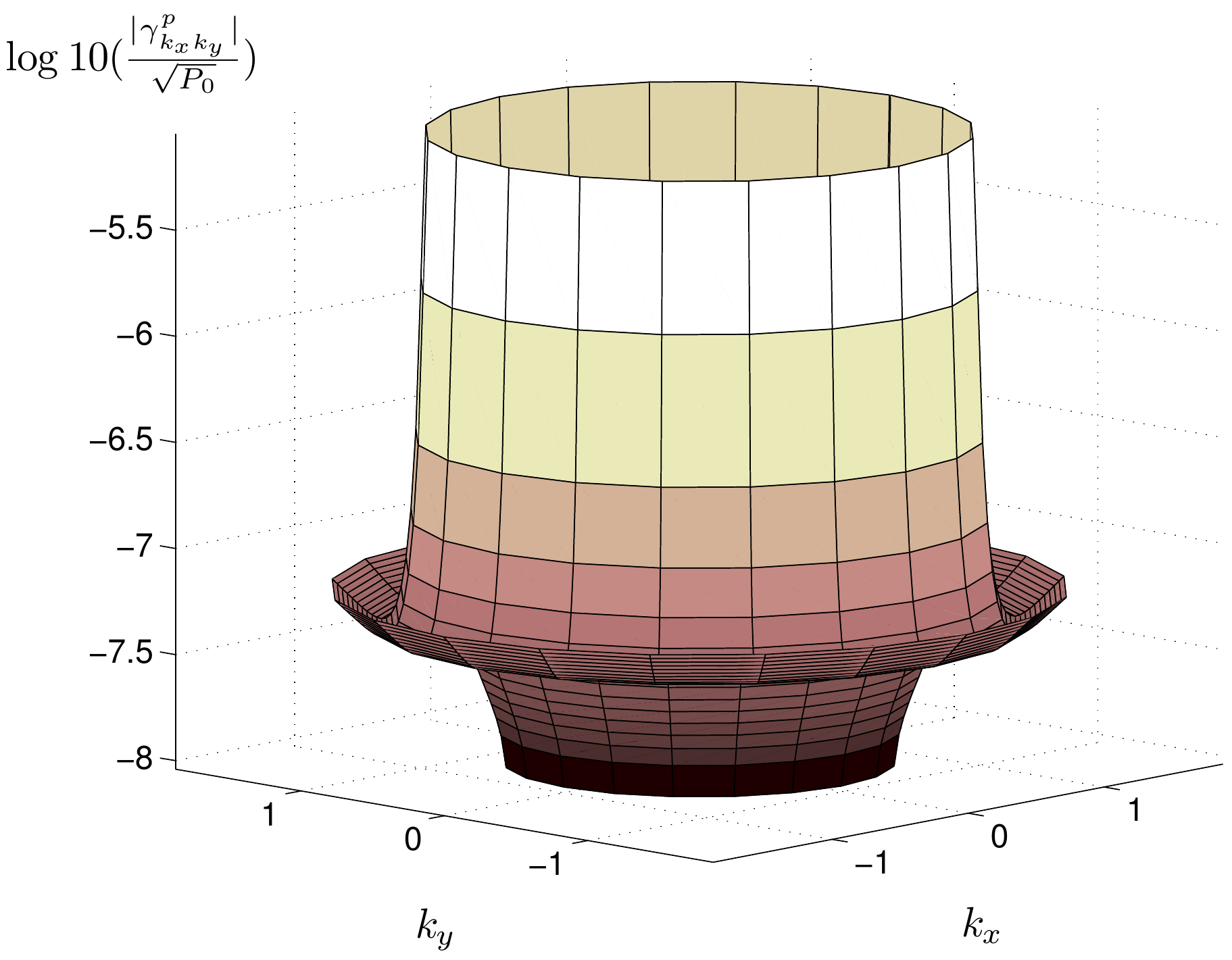}
			\put(90,70){\thesubfigure}
			\end{overpic}
			}
  \subfigure{%
            \label{fig1:D}
			\begin{overpic}[width=8cm,tics=10]{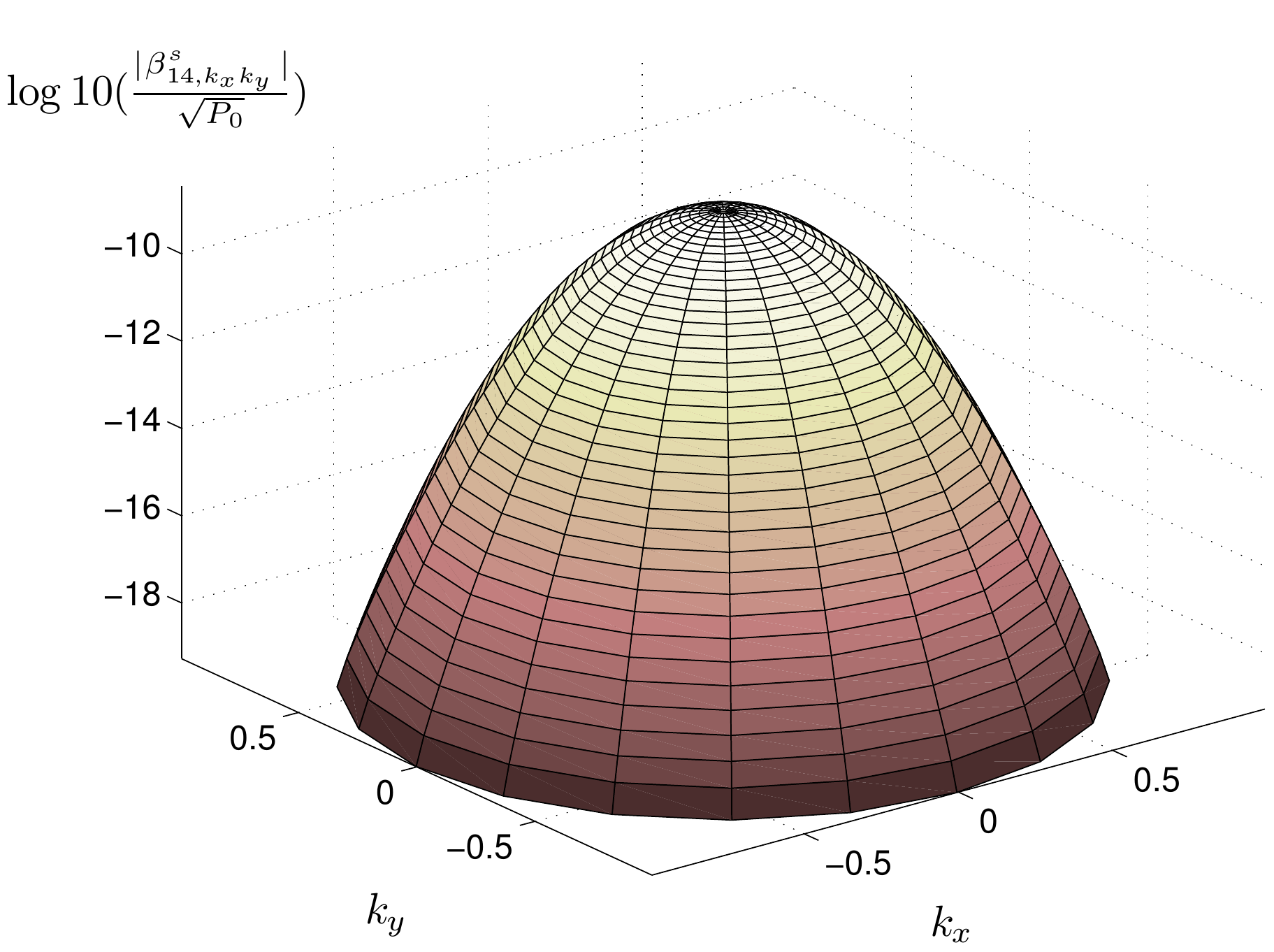}
			\put(85,65){\thesubfigure}
			\end{overpic}
			}
   \end{center}
\caption{\label{fig:fig1}(Color online) Logarithmic plots of the absolute value of the expansion coefficients $\gamma_{k_xk_y}^s$ (a) and $\gamma_{k_xk_y}^p$ (b), corresponding to the S and P polarization components of the angular spectrum of the scattered field and $\beta_{14,k_xk_y}^s$ (c), corresponding to the S polarization component of the angular spectrum of the direct field. The amplitudes are normalized to the total power ($P_0$) of the input beam in layer $l=1$, i.e. before going through the metal layer. The momentum axes are normalized to the vacuum wavenumber.}
\end{figure*}

Let us now analyze the results. Figs. \ref{fig1:S} and \ref{fig1:P} are three dimensional plots of $|\gamma^{s}_{k_xk_y}|$ and $|\gamma^{p}_{k_xk_y}|$ respectively. The clear resulting cylindrical symmetry of the scattered components is expected since structure, nanohole and illuminating beam have that symmetry. Note though, that none of the calculations in our method make use of the nanohole or beam symmetries: the method is independent of the shapes of the perturbation(s) and the illuminating beam, and its derivation does not contain any approximation. The strong presence of z-evanescent components illustrates the mechanism by which, through the volume integrals (\ref{eq:gamma}), the perturbation generates components with transversal wavevectors absent in the original external excitation. When the volume of the perturbation or its dielectric contrast tend to zero, so do the amplitudes of these new components. Fig. \ref{fig1:D} is the three dimensional plot of $|\beta^{s}_{14,k_xk_y}|$, the angular spectrum representation of the S component of the direct field distribution at the target layer as written in (\ref{eq:E0}) and (\ref{eq:beta}). The corresponding plot $|\beta^{p}_{14,k_xk_y}|$ is not included since it is practically identical. The direct field distribution is similar to the input Gaussian beam. It is mostly contained in the central region of the spectrum and decays rapidly with increasing normalized transversal wave vector $k_{\rho}=\sqrt{k_x^2+k_y^2}/k_0$. This similarity is due to the fact that the generalized transmission coefficients $\lsc{v}T_{14,k_xk_y}^\pm$ do not have big variations across the central $[k_x,k_y]$ region for this particular system. It is also apparent from Fig. \ref{fig:fig1} that there is much more scattered light than direct light. Indeed, a simple calculation reveals that the ratio of propagating power is about 250 times in favor of the scattered field.
\begin{figure*}[htbp]
\begin{center}
  \subfigure{%
            \label{fig2:S}
			\begin{overpic}[width=8cm,tics=10]{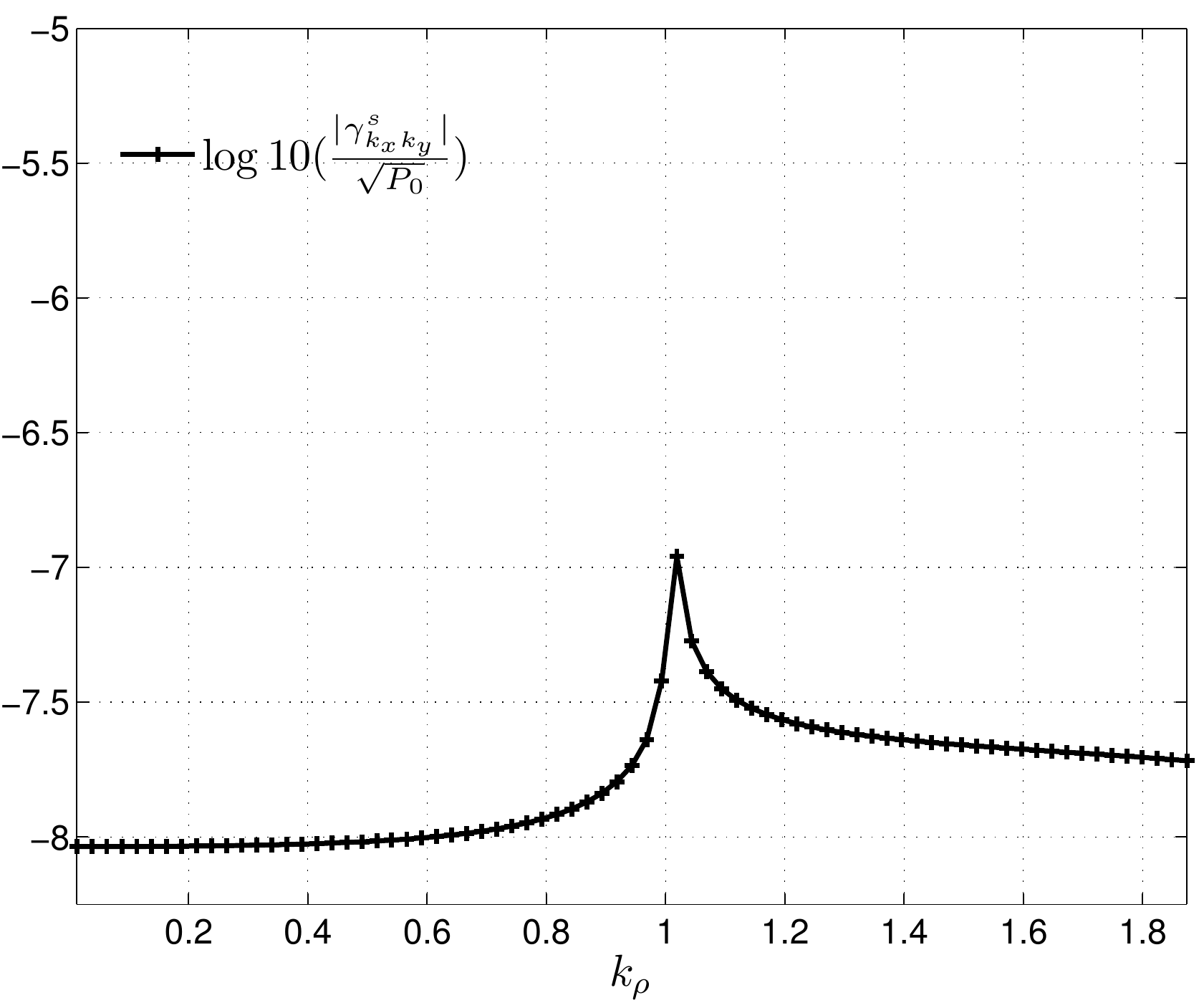}
			\put(90,75){\thesubfigure}
			\end{overpic}
			}
  \subfigure{%
            \label{fig2:P}
			\begin{overpic}[width=8cm,tics=10]{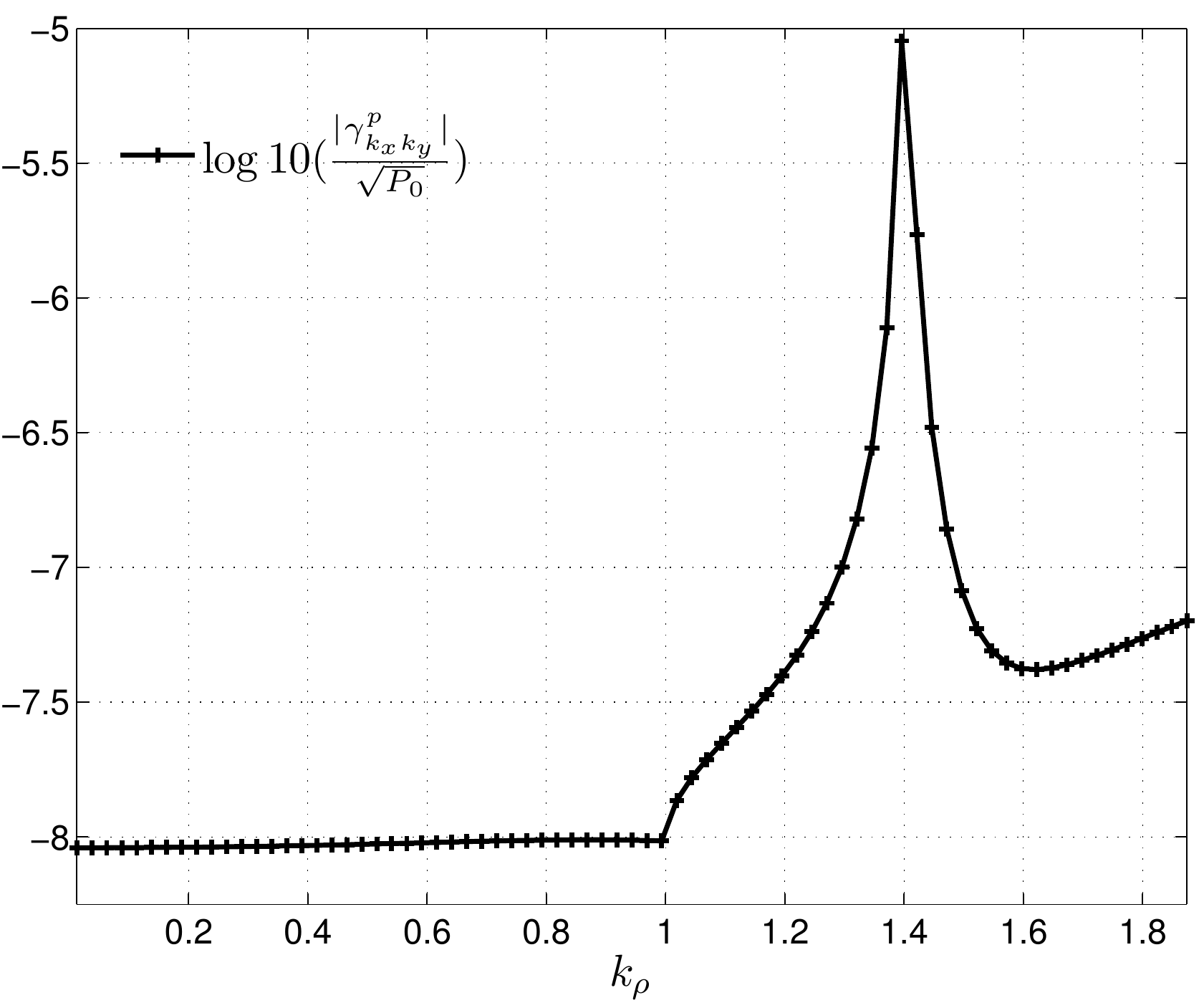}
			\put(90,75){\thesubfigure}
			\end{overpic}
			}
  \end{center}
\caption{\label{fig:radial} Logarithmic radial cut of (a) the S polarization components, $|\gamma^{s}_{k_xk_y}|$ and (b) the P polarization components $|\gamma^{p}_{k_xk_y}|$ of the scattered field, with respect to the normalized transversal wavevector $k_{\rho}=\sqrt{k_x^2+k_y^2}/k_0$.}
\end{figure*}

Due to the aforementioned cylindrical symmetry, radial cuts of the figures contain the same information as the three-dimensional (3D) plots. Figures \ref{fig2:S} and \ref{fig2:P} are radial cuts of the 3D angular spectrum coefficients as a function of $k_{\rho}$. The peaks in Figs. \ref{fig2:S} and \ref{fig2:P} correspond to normalized transversal wavevectors $k_{\rho}$ of 1.03 and 1.38. Both are evanescent in vacuum and their transversal wavevectors can be related to two resonant modes of the multilayered structure \cite{Li1987}. The P resonance position approximately matches the expected wavenumber of a surface plasmon at the gold-glass interface for the four layer system. It is interesting to see that a resonant S mode also appears. Even though it has a significantly smaller amplitude than the P mode, the difference in transversal wavevector means that the S mode field decay in the $z$ direction is much slower than that of the P mode. The existence of these S-polarized waves in multilayered systems and their role in the extraordinary optical transmission of light have already been discussed in the literature \cite{Moreno2006}.
\subsection{Other Field Functionals}
In this section, we have used different functions of the sets of coefficients $\gamma^s_{k_xk_y}$ and $\gamma^p_{k_xk_y}$ to expose some features of the angular spectrum representation of the scattered field of a nanohole under Gaussian illumination. It is worth highlighting that all the information contained in $\mathbf{E_{sc}}(\mathbf{r})$ is captured by those coefficients. Consequently, any functional of the field must be obtainable from them. Let us take for instance a common functional used in far field scattering analysis: the differential scattering cross section. Following closely \cite[sec. V]{Johansson2010}, we define it here as:

\begin{equation}
\label{eq:scs}
\frac{d\sigma}{d\Omega}=\eta |\mathbf{E_{sc}}(\mathbf{r})|^2,
\end{equation}
where point $r$ is supposed to be in the far field and $\eta$ depends on $|r|^2$ and the power of the illuminating beam. The differential scattering cross section is a function of the solid angle $\Omega$, and it is related to the far field electromagnetic power that the perturbation scatters through each solid angle. Alternatively, it is a function of two linear angles, polar and azimuthal. In the far field these angles can be computed as $\theta = \arcsin (k_{\rho})$ and $\phi = \arctan\left(\frac{k_y}{k_x}\right)$ respectively. 
The stationary phase approximation \cite[chap. 3.3]{Mandel1995} can be used to show that, at a point $r=[x,y,z]$ in the far field:
\begin{equation}
\label{eq:statphase}
|\mathbf{E_{sc}}(\mathbf{r})|^2=C|k_z|^2(|\gamma_{k_xk_y}^s|^2+|\gamma_{k_xk_y}^p|^2), 
\end{equation}
with $\frac{\sqrt{k_x^2+k_y^2}}{k_0}=\frac{\sqrt{x^2+y^2}}{\sqrt{x^2+y^2+z^2}}$ and $\frac{k_y}{k_x}=\frac{y}{x}$, and $C$ is constant in the variables of interest.

In our case, due to the cylindrical symmetry of the results, (\ref{eq:scs}) will only depend on one of the angles, the polar angle $\theta=\arcsin (k_{\rho})$. To include only propagating components, it suffices to restrict $k_{\rho}<1$. Finally, to obtain $\frac{d\sigma}{d\theta}$, rather than $\frac{d\sigma}{d\Omega}$, we make use of the fact that in our coordinate system $d\Omega=d\phi d\theta \cos(\theta)$.

\begin{figure}[htbp]
\includegraphics[width=8cm]{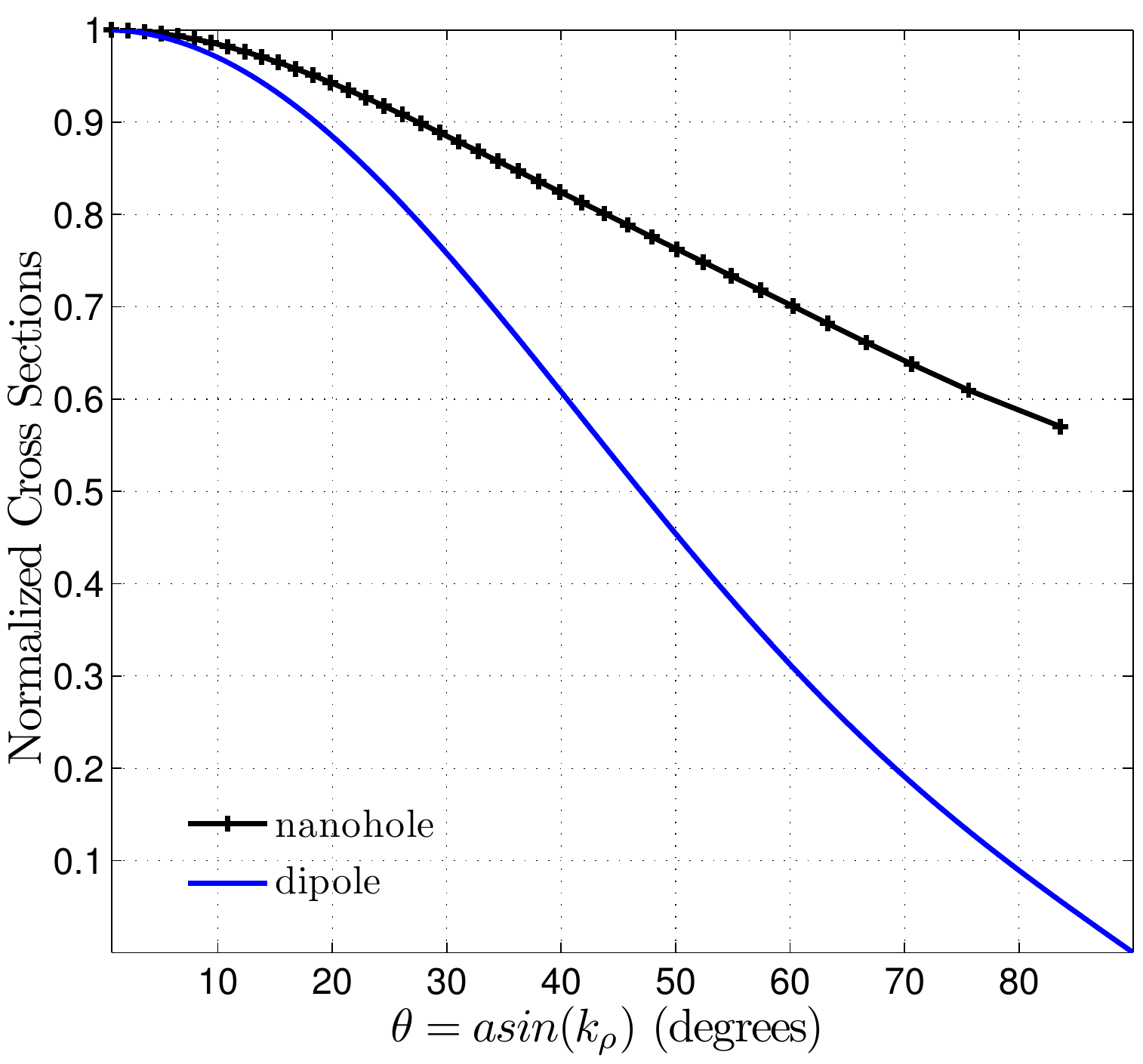}
\caption{\label{fig3:3}(Color online) Differential scattering cross sections of the nanohole (cross-solid black line) and a $\mathbf{\hat{x}}-i\mathbf{\hat{y}}$ oriented dipole (blue solid line). Both lines are normalized to their values at $k_{\rho}=0$.}
\end{figure}

For reference purposes, we provide the differential scattering cross section of the nanohole in Fig. \ref{fig3:3} together with that of an infinitesimal electric dipole. Since our illuminating beam has right circular polarization, it is appropriate to choose the orientation of the dipole to match it. The dipole is oriented along the $\mathbf{\hat{x}}-i\mathbf{\hat{y}}$ direction. Its angular spectrum can be computed following \cite[chap. 2.12.1]{Novotny2006}, and the previously outlined steps can be used again to obtain its differential scattering cross section.

It can be noticed that the behavior of the scattered light from the nanohole is rather different than that of a single dipole. This is expected due to two main reasons: (a) on the one hand, in our case the size of the nanohole is only a fraction of the wavelength of radiation, meaning that we expect our radiation to be multipolar, and (b) on the other hand, the coupling to the surfaces modes greatly affects the radiation from the nanohole, as can be seen in Figs. \ref{fig:fig1} and \ref{fig:radial}.

\section{Conclusion}
In summary, we have presented a method to obtain a plane wave expansion of the field scattered by a perturbation present in a planar multilayered structure under general illumination. Once the field $\mathbf{E}(\mathbf{r'})$ inside the perturbation is known, this method provides a very compact expansion of the scattered field, which allows for a straightforward analysis.

The technique allows access to all the information contained in the scattered field by providing the amplitude and phase of the complex coefficients multiplying each plane wave for both orthogonal polarizations. This complete information can be easily manipulated to obtain other more common field dependent functions like the far field radiation diagrams and differential scattering cross sections, relevant when studying the far field properties of the light scattered by perturbations in a planar multilayered system. With respect to near-field applications like surface-enhanced Raman scattering or near-field microscopy, the technique should prove useful since it provides all the information about the evanescent components of the field, which are critical for these applications.

In this article, the method has been applied to the problem of a nanohole in a thin gold film illuminated by a realistic Gaussian beam. The exact angular spectrum of its scattered field has been reported.\footnote{During the review process of our article, the radiation diagram of a similar system has been published in \cite{Bordo2011}. The author uses an elegant analytical technique which is valid for single cylindrical dielectric perturbations in a multilayered system. The radiation diagram, related to the propagating part of the angular spectrum, is obtained under some approximations: optically thick metal layer, far-field scattering and elongated nanohole (hole diameter is much smaller than its longitudinal dimension).}. The diagrams present two distinct peaks in the evanescent region, one for each polarization. Both peaks can be traced back to resonant modes of the multilayered structure.

Our technique can be easily extended to problems with different geometries. From Eq. (\ref{eq:GT}) on, our analysis is specific to planar geometries stratified in the $\mathbf{\hat{z}}$ direction. Nevertheless, Ref. \cite[sec. 6]{Tan1998} contains the Green's function decomposition of spherical and cylindrical multilayered structures as well, with $\hat{r}$ and $\hat{\rho}$ stratification directions. The steps in the previous derivation can be repeated to obtain decompositions of scattered fields in cylindrical and spherical vector wave functions. In the case of spherical geometry the expressions will be formally identical to those obtained in Sec. \ref{sec:method}. For the cylindrical geometry, according to \cite{Xiang1996} and \cite{Tan1998}, there will generally be cross terms in the expansion of the GT of the kind $\mathbf{M}(\mathbf{r})\otimes \mathbf{N}(\mathbf{r'})$ and $\mathbf{N}(\mathbf{r})\otimes \mathbf{M}(\mathbf{r'})$ (seen explicitly in \cite{Xiang1996}). This is not a blocking point. Since the separate dependence in $r$ and $\mathbf{r'}$ is maintained, our derivation can still be carried out by introducing more terms besides $e$ and $f$ if necessary. 

We advance that this method can have an important impact in elucidating the different mechanisms of electromagnetic interaction between different features of nanostructures. The flexibility of extending this method to other geometries will prove very useful in a large number of applications. 
\\
\\
\section{Acknowledgements} 
This research was supported under the Australian Research Council's Discovery Projects funding scheme (Project No. DP110103697).
    

\end{document}